\documentclass[12pt]{article}
  \usepackage{amsfonts}
  \usepackage{amsmath}
\usepackage{amssymb}
\usepackage{amscd}
\usepackage{graphicx}
\begin{document}

~~
\bigskip
\bigskip
\begin{center}
{\Large {\bf{{{The energy-momentum conservation law in two-particle system for twist-deformed Galilei Hopf algebras}}}}}
\end{center}
\bigskip
\bigskip
\bigskip
\begin{center}
{{\large ${\rm {Marcin\;Daszkiewicz}}$}}
\end{center}
\bigskip
\begin{center}
\bigskip

{ ${\rm{Institute\; of\; Theoretical\; Physics}}$}

{ ${\rm{ University\; of\; Wroclaw\; pl.\; Maxa\; Borna\; 9,\;
50-206\; Wroclaw,\; Poland}}$}

{ ${\rm{ e-mail:\; marcin@ift.uni.wroc.pl}}$}

\end{center}
\bigskip
\bigskip
\bigskip
\bigskip
\bigskip
\bigskip
\bigskip
\bigskip
\bigskip
\begin{abstract}
In this article we discus the energy-momentum conservation principle for two-particle system in the case of canonically and Lie-algebraically twist-deformed
Galilei Hopf algebra. Particularly, we provide consistent with the coproducts energy and momentum addition law as well as its symmetric with respect the
exchange of particles counterpart. Besides, we show that the vanishing of total fourmomentum for two Lie-algebraically deformed kinematical models leads to the discret
values of energies and momenta only in the case of the symmetrized addition rules.
\end{abstract}
\bigskip
\bigskip
\bigskip
\bigskip
\eject

\section{{{Introduction}}}

The suggestion to use noncommutative coordinates goes back to
Heisenberg and was firstly  formalized by Snyder in \cite{snyder}.
Recently, there were also found formal  arguments based mainly  on
Quantum Gravity \cite{2}, \cite{2a} and String Theory models
\cite{recent}, \cite{string1}, indicating that space-time at the Planck
scale  should be noncommutative, i.e., it should  have a quantum
nature. Consequently, there appeared a lot of papers dealing with
noncommutative classical and quantum  mechanics (see e.g.
\cite{mech}, \cite{qm}) as well as with field theoretical models
(see e.g. \cite{prefield}, \cite{field}), in which  the quantum
space-time is employed.

In accordance with the Hopf-algebraic classification of all
deformations of relativistic \cite{class1} and nonrelativistic
\cite{class2} symmetries, one can distinguish three basic types
of space-time noncommutativity (see also \cite{nnh} for details):\\
\\
{ \bf 1)} Canonical ($\theta^{\mu\nu}$-deformed) type of quantum space \cite{oeckl}-\cite{dasz1}
\begin{equation}
[\;{ x}_{\mu},{ x}_{\nu}\;] = i\theta_{\mu\nu}\;, \label{noncomm}
\end{equation}
\\
{ \bf 2)} Lie-algebraic modification of classical space-time \cite{dasz1}-\cite{lie1}
\begin{equation}
[\;{ x}_{\mu},{ x}_{\nu}\;] = i\theta_{\mu\nu}^{\rho}{ x}_{\rho}\;,
\label{noncomm1}
\end{equation}
and\\
\\
{ \bf 3)} Quadratic deformation of Minkowski and Galilei  spaces \cite{dasz1}, \cite{lie1}-\cite{paolo}
\begin{equation}
[\;{ x}_{\mu},{ x}_{\nu}\;] = i\theta_{\mu\nu}^{\rho\tau}{
x}_{\rho}{ x}_{\tau}\;, \label{noncomm2}
\end{equation}
with coefficients $\theta_{\mu\nu}$, $\theta_{\mu\nu}^{\rho}$ and  $\theta_{\mu\nu}^{\rho\tau}$ being constants.\\
\\
Moreover, it has been demonstrated in \cite{nnh}, that in the case of the
so-called N-enlarged Newton-Hooke Hopf algebras
$\,{\mathcal U}^{(N)}_0({ NH}_{\pm})$ the twist deformation
provides the new  space-time noncommutativity of the
form\footnote{$x_0 = ct$.},\footnote{ The discussed space-times have been  defined as the quantum
representation spaces, so-called Hopf modules (see e.g. \cite{oeckl}, \cite{chi}), for the quantum N-enlarged
Newton-Hooke Hopf algebras.}
\begin{equation}
{ \bf 4)}\;\;\;\;\;\;\;\;\;[\;t,{ x}_{i}\;] = 0\;\;\;,\;\;\; [\;{ x}_{i},{ x}_{j}\;] = 
if_{\pm}\left(\frac{t}{\tau}\right)\theta_{ij}(x)
\;, \label{nhspace}
\end{equation}
with time-dependent  functions
$$f_+\left(\frac{t}{\tau}\right) =
f\left(\sinh\left(\frac{t}{\tau}\right),\cosh\left(\frac{t}{\tau}\right)\right)\;\;\;,\;\;\;
f_-\left(\frac{t}{\tau}\right) =
f\left(\sin\left(\frac{t}{\tau}\right),\cos\left(\frac{t}{\tau}\right)\right)\;,$$
$\theta_{ij}(x) \sim \theta_{ij} = {\rm const}$ or
$\theta_{ij}(x) \sim \theta_{ij}^{k}x_k$ and  $\tau$ denoting the time scale parameter
 -  the cosmological constant. Besides, it should be  noted that the all above quantum spaces {\bf 1)}, { \bf 2)} and { \bf 3)}
can be obtained  by the proper contraction limit  of the commutation relations { \bf 4)}\footnote{Such a result indicates that the twisted N-enlarged Newton-Hooke Hopf algebra plays a role of the most general type of quantum group deformation at nonrelativistic level.}.

It is commonly known that the above mentioned Poincare and Galilei Hopf algebras are given by the algebraic as well as by the coalgebraic sector respectively. Since the first of them defines
the commutation relationes for the generators, the second one introduces particularly the addition rules of two momenta and energies. Usually, such a constructed composition law remains nonsymmetric with respect the exchange of particles \cite{comp1}-\cite{compn} and the different solutions of
the problem have been proposed in papers \cite{comp1}, \cite{comp2} and \cite{compn}\footnote{The symmetrization problem has been mainly considered in the case of $\kappa$-Poincare Hopf structure
${\cal U}_{\kappa}({\cal P})$.}.

In this article we provide consistently with coproduct the fourmomentum addition laws for two canonically and three Lie-algebraically twist-deformed Galilei Hopf algebras \cite{dasz1}.
We show that after the proper symmetrization only two of them (for ${\cal U}_{\kappa_1}({\cal G})$ and ${\cal U}_{\kappa_2}({\cal G})$ quantum groups) are still deformed while the third
one becomes classical. Further, we demonstrate that the corresponding two-particle energy-momentum conservation principle with vanishing conserved quantity $\bar{p}_{\rm {tot}}$ and
$E_{\rm {tot}}$ leads to the discret values of momenta and energies only in the case of the symmetrized addition rules {\bf i)}, {\bf ii)} and {\bf iii)}\footnote{Before symmetrization the condition $\bar{p}_{\rm {tot}}=0$ and
$E_{\rm {tot}}=0$ gives the classical solution $\bar{p}_1=-\bar{p}_2$ and $E_1=-E_2$ respectively.}.

The paper is organized as follows. In second Section we remaind the canonically-deformed Galilei Hopf algebras proposed in \cite{dasz1}. The third Section is devoted to the corresponding
fourmomentum conservation principle. In the fourth Section we recall the Lie-algebraically nonrelativistic Hopf structure while the proper conservation rules are provided and analyzed
in fifth Section. The final remarks are discussed in the last Section.

\section{Canonical twist-deformations of Galilei Hopf algebra}

The two canonically deformed Galilei Hopf algebras ${\cal U}_{\theta_{ij}}({\cal G})$ and ${\cal U}_{\theta_{0i}}({\cal G})$ have been provided in article \cite{dasz1}
by the proper contractions of their relativistic counterparts. They are given by the classical algebraic sector
\begin{eqnarray}
&&\left[\, K_{ij},K_{kl}\,\right] =i\left( \delta
_{il}\,K_{jk}-\delta
_{jl}\,K_{ik}+\delta _{jk}K_{il}-\delta _{ik}K_{jl}\right) \;,  \label{ff} \\
&~~&  \cr &&\left[\, K_{ij},V_{k}\,\right] =i\left( \delta
_{jk}\,V_i-\delta _{ik}\,V_j\right)\;\; \;, \;\;\;\left[
\,K_{ij},\Pi_{\rho }\right] =i\left( \eta _{j \rho }\,\Pi_{i }-\eta
_{i\rho }\,\Pi_{j }\right) \;, \label{nnn}
\\
&~~&  \cr &&\left[ \,V_i,V_j\,\right] = \left[ \,V_i,\Pi_{j
}\,\right] =0\;\;\;,\;\;\;\left[ \,V_i,\Pi_{0 }\,\right]
=-i\Pi_i\;\;\;,\;\;\;\left[ \,\Pi_{\rho },\Pi_{\sigma }\,\right] =
0\;,\label{ff1}
\end{eqnarray}
where $K_{ij}$, $\Pi_0$, $\Pi_i$ and $V_i$ can be identified with rotation, time translation, momentum and boost operators as well as by the following
twisted coproducts
\begin{eqnarray}
&&\Delta_{\theta^{ij}}(\Pi_\mu)=\Delta_0(\Pi_\mu)\;\;\;,\;\;\;
\Delta _{\theta^{ij} }(V_i) =\Delta _{0}(V_i)\;, \label{dlww3v}\\
&~~~&  \cr \Delta _{\theta^{ij} }(K_{ij})
&=&\Delta _{0}(K_{ij})-%
\theta ^{k l }[(\delta_{k i}\Pi_{j }-\delta_{k j
}\,\Pi_{i})\otimes \Pi_{l }\nonumber\\
&&\qquad\qquad\qquad\qquad\qquad+\Pi_{k}\otimes (\delta_{l
i}\Pi_{j}-\delta_{l j}\Pi_{i})]\;,\label{zadruzny}
\end{eqnarray}
and
\begin{eqnarray}
\Delta_{\theta^{0i}}(\Pi_\mu)&=&\Delta _0(\Pi_\mu)\;,\label{zcoppy1}\\
 &~~&  \cr
\Delta_{\theta^{0i}}(K_{ij})&=&\Delta_0(K_{ij})
- {{\theta}^{0k}}\Pi_0
\wedge\left(\delta_{ki}\Pi_j - \delta_{kj}\Pi_i\right)\;,\\
 &~~&  \cr
\Delta_{\theta^{0i}}(V_i)&=&\Delta_0(V_i) -
{{\theta}^{0k}}\Pi_i \wedge
\Pi_k\;,\label{zcoppy100}
\end{eqnarray}
respectively. Besides, in the case of ${\cal U}_{\theta_{ij}}({\cal G})$ Hopf structure the corresponding quantum space-time is given by
\begin{eqnarray}
\left[\;{t},{x}_i\;\right] = 0\;\;\;,\;\;\;\left[\;{x}_i,{x}_j\;\right] = i{\theta_{ij}}\;, \label{eq5}
\end{eqnarray}
while for  ${\cal U}_{\theta_{0i}}({\cal G})$ it looks as follows
\begin{eqnarray}
\left[\;{t},{x}_i\;\right] = i\theta_{0i}\;\;\;,\;\;\;\left[\;{x}_i,{x}_j\;\right] = 0\;. \label{equs5}
\end{eqnarray}
Of course, for deformation parameters $\theta_{ij}$ and $\theta_{0i}$ approaching zero the above relations become classical.

\section{Canonically deformed energy-momentum conservation law}

Let us turn to the momentum addition rules corresponding to the $\Delta(P)$-coproducts (\ref{dlww3v}) and (\ref{zcoppy1}). Due to the fact that both of them are primitive,
for momenta $\bar{p}_1=[\;p_{11},p_{12},p_{13}\;]$ and $\bar{p}_2=[\;p_{21},p_{22},p_{23}\;]$ as well as for energies $E_1$ and $E_2$, we have
\begin{eqnarray}
\bar{p}_1 +_A \bar{p}_2 &=& \bar{p}_3\;\;;\;\;A=\theta_{ij}\; {\rm or}\; A=\theta_{i0}\;, \label{conserv1}\\
E_1 +_A E_2 &=& E_3\;, \label{conserv2}
\end{eqnarray}
with
\begin{eqnarray}
\bar{p}_3 &=& [\;p_{11}+p_{21},p_{12}+p_{22},p_{13}+p_{23}\;]\;, \label{conserva1}\\
E_3 &=& E_1 + E_2 \;, \label{conserva2}
\end{eqnarray}
in the case of ${\cal U}_{\theta_{ij}}({\cal G})$ and  ${\cal U}_{\theta_{i0}}({\cal G})$ Hopf structures. It means that for both algebras the addition law remains undeformed and the energy-momentum
conservation principle for two-particle system takes the standard form
\begin{eqnarray}
\bar{p}_1 +_A \bar{p}_2 &=& \bar{p}_1 + \bar{p}_2\; = \;\bar{p}_{{\rm tot}} \;= \;{\rm const.}\;, \label{law1}\\
E_1 +_A E_2 &=& E_1 + E_2 \;= \;E_{{\rm tot}} \;= \;{\rm const.}\;. \label{law2}
\end{eqnarray}
Of course, for $\bar{p}_{{\rm tot}}$ and $E_{{\rm tot}}$ equal to zero we get
\begin{eqnarray}
\bar{p}_1 = - \bar{p}_2 \;, \label{forem1}
\end{eqnarray}
and
\begin{eqnarray}
E_1 = - E_2\;, \label{forem2}
\end{eqnarray}
respectively.

\section{Twisted Lie-algebraically deformed Galilei Hopf st-ructures}

The three Lie-algebraically twist-deformed Galilei Hopf structures ${\cal U}_{\kappa_1}({\cal G})$, ${\cal U}_{\kappa_2}({\cal G})$ and ${\cal U}_{\kappa_3}({\cal G})$ have been
introduced in article \cite{dasz1} as well. Their algebraic sectors remain classical (see formulas (\ref{ff})-(\ref{ff1})) while the coproducts are given by\footnote{The indexes $k$, $l$, $\gamma$ are fixed, spatial and different.},\footnote{$a\wedge b = a\otimes b - b\otimes a\;,\; a\perp b = a\otimes b+b\otimes a$.},\footnote{$\psi_\lambda =\eta_{\nu
\lambda }\eta_{\beta \mu}-\eta_{\mu \lambda }\eta_{\beta
\nu}\;,\;\chi_\lambda =\eta_{\nu \lambda }\eta_{\alpha
\mu}-\eta_{\mu \lambda }\eta_{\alpha \nu}$.}
\begin{eqnarray}
 \Delta_{\kappa_1}(\Pi_0)&=&\Delta _0(\Pi_0)\;,\label{ffdlww2.2}\\
&~~&  \cr \Delta_{\kappa_1}(\Pi_i)&=&\Delta _0(\Pi_i)+\sin\left( \frac{1}{\kappa_1}
\Pi_\gamma \right)\wedge
\left(\delta_{k i}\Pi_l -\delta_{l i}\Pi_k \right)\label{america1}\\
&+&\left[\cos\left(\frac{1}{\kappa_1}  \Pi_\gamma \right)-1\right]\perp \left(\delta_{k i}\Pi_k
+\delta_{l i}\Pi_l \right)\;, \nonumber
\end{eqnarray}
\begin{eqnarray}
~~ \Delta_{\kappa_1}(K_{ij})&=&\Delta_0(K_{ij})+K_{k l }\wedge \frac{1}{\kappa_1}
\left(\delta_{i
\gamma }\Pi_j-\delta_{j \gamma }\Pi_i\right)\nonumber\\
&+&i\left[K_{ij},K_{k l }\right]\wedge
\sin\left(\frac{1}{\kappa_1} \Pi_\gamma \right) \notag \\
&+&\left[\left[%
K_{ij},K_{k l }\right],K_{k l }\right]\perp
\left[\cos\left(\frac{1}{\kappa_1}  \Pi_\gamma  \right)-1\right]  \label{ffdlww2.22} \\
&+&K_{k l }\sin\left(\frac{1}{\kappa_1} \Pi_\gamma \right)\perp
\frac{1}{\kappa_1} \left(\psi_\gamma \Pi_k -\chi_\gamma \Pi_l \right) \notag \\
&+&\frac{1}{\kappa_1} \left(\psi_\gamma \Pi_l +\chi_\gamma \Pi_k \right)\wedge K_{k
l }\left[\cos\left(\frac{1}{\kappa_1} \Pi_\gamma \right)-1\right]\;,
  \notag
\end{eqnarray}
\begin{eqnarray}
\Delta_{\kappa_1}(V_i)&=&\Delta_0(V_i) +i\left[V_i,K_{k l }\right]\wedge
\sin\left(\frac{1}{\kappa_1} \Pi_\gamma \right)
\label{sfifdlww2.22}\\
&+&\left[\left[%
V_i,K_{k l }\right],K_{k l }\right]\perp \left[\cos\left(\frac{1}{\kappa_1} \Pi_\gamma
\right)-1\right]\;,  \nonumber
\end{eqnarray}
in case of the first quantum group
\begin{eqnarray}
 \Delta_{\kappa_2}(\Pi_0)&=&\Delta _0(\Pi_0)\;,\label{sffdlww2.2}\\
&~~&  \cr \Delta_{\kappa_2}(\Pi_i)&=&\Delta _0(\Pi_i)+\sin\left( \frac{1}{\kappa_2}
\Pi_0 \right)\wedge
\left(\delta_{k i}\Pi_l -\delta_{l i}\Pi_k \right)\label{america2}\\
&+&\left[\cos\left(\frac{1}{\kappa_2}  \Pi_0 \right)-1\right]\perp \left(\delta_{k i}\Pi_k
+\delta_{l i}\Pi_l \right)\;, \nonumber
\end{eqnarray}
\begin{eqnarray}
~~ \Delta_{\kappa_2}(K_{ij})&=&\Delta_0(K_{ij})+K_{k l }\wedge \frac{1}{\kappa_2}
\left(\delta_{i
0 }\Pi_j-\delta_{j 0 }\Pi_i\right)\nonumber\\
&+&i\left[K_{ij},K_{k l }\right]\wedge
\sin\left(\frac{1}{\kappa_2} \Pi_0 \right) \notag \\
&+&\left[\left[%
K_{ij},K_{k l }\right],K_{k l }\right]\perp
\left[\cos\left(\frac{1}{\kappa_2}  \Pi_0  \right)-1\right]  \label{sffdlww2.22} \\
&+&K_{k l }\sin\left(\frac{1}{\kappa_2} \Pi_0 \right)\perp
\frac{1}{\kappa_2} \left(\psi_0 \Pi_k -\chi_0 \Pi_l \right) \notag \\
&+&\frac{1}{\kappa_2} \left(\psi_0 \Pi_l +\chi_0 \Pi_k \right)\wedge K_{k
l }\left[\cos\left(\frac{1}{\kappa_2} \Pi_0 \right)-1\right]\;,
  \notag
\end{eqnarray}
\begin{eqnarray}
\Delta_{\kappa_2}(V_i)&=&\Delta_0(V_i)+\frac{1}{{{\kappa_2}}}K_{k
l}\wedge
 \Pi_i+i\left[V_i,K_{k l }\right]\wedge
\sin\left(\frac{1}{{{\kappa_2}}} \Pi_0 \right) \notag \\
&+&\left[\left[%
V_i,K_{k l }\right],K_{k l }\right]\perp
\left[\cos\left(\frac{1}{{{\kappa_2}}}  \Pi_0  \right)-1\right]  \label{copp2} \\
&+&K_{k l }\sin\left(\frac{1}{{{\kappa_2}}} \Pi_0 \right)\perp
\frac{1}{{{\kappa_2}}} \left(\delta_{k i}\Pi_l - \delta_{l i} \Pi_k  \right) \notag \\
&-&\frac{1}{{{\kappa_2}}} \left(\delta_{k i}\Pi_k+\delta_{l i}
\Pi_l \right)\wedge K_{k l }\left[\cos\left(\frac{1}{{{\kappa_2}}} \Pi_0
\right)-1\right]\;,\notag
\end{eqnarray}
for the second Hopf algebra and
\begin{eqnarray}
 \Delta_{{\kappa_3}}(\Pi_0)&=&\Delta _0(\Pi_0) +
\frac{1}{{{\kappa_3}}} \Pi_l \wedge \Pi_k\;,\label{coppy1}\\
 &~~&  \cr
\Delta_{{\kappa_3}}(\Pi_i)&=&\Delta
_0(\Pi_i)\;\;\;,\;\;\;\Delta_{{\kappa_3}}(V_i)=\Delta_0(V_i)\;,\label{cop0}\\
 &~~&  \cr
\Delta_{{\kappa_3}}(K_{ij})&=&\Delta_0(K_{ij})+
\frac{i}{{{\kappa_3}}}\left[K_{ij},V_k\right]\wedge \Pi_l +
\frac{1}{{{\kappa_3}}}V_k \wedge(\delta_{il}\Pi_j
-\delta_{jl}\Pi_i)  \;, \label{coppy100}
\end{eqnarray}
for the third, ${\cal U}_{\kappa_3}({\cal G})$ Hopf structure. One can also check that that the corresponding quantum space-times look as follows
\begin{equation}
[\, x_{i },x_{j }\,] = \frac{i}{\kappa_1} \delta_{\gamma j}(
 \delta _{k i }x_{l }- \delta_{l i }x_{k }) +\frac{i}{\kappa_1} \delta_{\gamma
i}(\delta _{ l j }x_{k } - \delta _{k j }x_{l } )
\;\;\;,\;\;\;[\,t,x_i\,] =  0\;, \label{ssstar}
\end{equation}
\begin{equation}
[\, t,x_{i }\,]= \frac{i}{{\kappa_2}} (
\delta _{ li }x_{k }-\delta _{k i }x_{l } )\;\;\;,\;\;\;[\,x_{i
},x_j\,] =  0 \;, \label{sistar}
\end{equation}
and
\begin{equation}
[\, x_{i },x_{j }\,]=
\frac{i}{{\kappa_3}}t(\delta _{l i }\delta_{k j }-\delta _{
ki }\delta _{l j
})\;\;\;,\;\;\;[\,t,x_i\,] = 0 \;.
\label{ysesstar}
\end{equation}
respectively. Obviously, for all deformation parameters $\kappa_1$, $\kappa_2$ and $\kappa_3$ running to infinity the above relations become commutative.

\section{Lie-algebraically deformed energy-momentum conservation law}

Due to the $\Delta(P)$-coproducts (\ref{ffdlww2.2}), (\ref{america1}), (\ref{sffdlww2.2}), (\ref{america2}), (\ref{coppy1}) and (\ref{cop0}) the two-particle energy-momentum addition rules for
$\bar{p}_1=[\;p_{1\gamma},p_{1k},p_{1l}\;]$ and $\bar{p}_2=[\;p_{2\gamma},p_{2k},p_{2l}\;]$ as well as for energies $E_1$ and $E_2$ take the form
\begin{eqnarray}
\bar{p}_1 +_{\kappa_i} \bar{p}_2 &=& \bar{p}_3\;\;;\;\;i=1,2,3\;, \label{lieconserv1}\\
E_1 +_{\kappa_i} E_2 &=& E_3\;, \label{lieconserv2}
\end{eqnarray}
with
\begin{eqnarray}
p_{3\gamma} &=& p_{1\gamma} +p_{2\gamma}\;, \label{p1}\\
p_{3k} &=& p_{1k} + p_{2k} +\sin\left( \frac{p_{1\gamma}}{\kappa_1}\right)p_{2l} -\sin\left( \frac{p_{2\gamma}}{\kappa_1}\right) p_{1l} \;+\nonumber\\
&~~&~~~~~~~~~+\;\left[\;\cos\left(\frac{ p_{1\gamma}}{\kappa_1}  \right)-1\right]p_{2k}\; +  \label{p2}\\
&~~&~~~~~~~~~+\;\left[\;\cos\left(\frac{ p_{2\gamma}}{\kappa_1}  \right)-1\;\right]p_{1k}\;,\nonumber
\end{eqnarray}
\begin{eqnarray}
p_{3l} &=& p_{1l} + p_{2l} - \sin\left( \frac{ p_{1\gamma}}{\kappa_1}\right)p_{2k} + \sin\left( \frac{ p_{2\gamma}}{\kappa_1}\right)p_{1k} \;+\nonumber\\
&~~&~~~~~~~~~+\;\left[\;\cos\left(\frac{ p_{1\gamma}}{\kappa_1}  \right)-1\;\right]p_{2l} \;+\label{p3}\\
&~~&~~~~~~~~~+\;\left[\;\cos\left(\frac{ p_{2\gamma}}{\kappa_1}  \right)-1\;\right]p_{1l}\;, \nonumber\\
E_3 &=& E_1 \;+\; E_2\;, \label{en}
\end{eqnarray}
in case of the first Hopf algebra
\begin{eqnarray}
p_{3\gamma} &=& p_{1\gamma} + p_{2\gamma}\;, \label{sp1}
\end{eqnarray}
\begin{eqnarray}
p_{3k} &=& p_{1k}+p_{2k}+ \sin\left( \frac{E_1}{\kappa_2}\right)p_{2l}-\sin\left( \frac{E_2}{\kappa_2}\right)p_{1l} \;+\nonumber\\
&~~&~~~~~~~~~+\;\left[\;\cos\left(\frac{E_1}{\kappa_2} \right)-1\;\right]p_{2k}\;+ \label{sp2}\\
&~~&~~~~~~~~~+\;\left[\;\cos\left(\frac{E_2}{\kappa_2} \right)-1\;\right]p_{1k}\;,\nonumber
\end{eqnarray}
\begin{eqnarray}
p_{3l} &=& p_{1l} \;+\; p_{2l}\; - \;\sin\left( \frac{E_1}{\kappa_2}\right)p_{2k}\; + \;p_{1k}\sin\left( \frac{E_2}{\kappa_2}\right) \;+\nonumber\\
&+&\left[\;\cos\left(\frac{E_1}{\kappa_2} \right)-1\;\right]p_{2l}\; + \;p_{1l}\left[\;\cos\left(\frac{E_2}{\kappa_2} \right)-1\;\right]
\;, \label{sp3}\\
E_3 &=& E_1\; +\; E_2\;, \label{sen}
\end{eqnarray}
for the second quantum group and
\begin{eqnarray}
&&p_{3\gamma} \;=\; p_{1\gamma} + p_{2\gamma}\;\;,\;\;
p_{3k} \;=\; p_{1k} + p_{2k}\;\;,\;\;
p_{3l} \;=\; p_{1l} + p_{2l}\;, \label{tp3}\\
&&E_3 \;=\; E_1 \;+ \;E_2 \;+\; \frac{1}{\kappa_3}\left(p_{1l}p_{2k}\; -\; p_{1k}p_{2l}\right)\;, \label{ten}
\end{eqnarray}
for the third Hopf structure. Consequently, the components (\ref{p1})-(\ref{ten}) become nonsymmetric with respect the exchange of indexes 1 and 2. In order
to improve the problem we modify the above rules in the most simple and natural way as follows\footnote{The terms of laws (\ref{p2}), (\ref{p3}) and (\ref{sp2}), (\ref{sp3})
including cosinus function are together symmetric with respect the exchange of indexes 1 and 2. Hence, they remain untouched by our procedure. However, the terms with sinus
function are together antysymmetric. It is easy to see that their following symmetrization (for example in the case of the first Hopf algebra)
$$
\sin\left( \frac{p_{1\gamma}}{\kappa_1}\right)p_{2l} - p_{1l}\sin\left( \frac{p_{2\gamma}}{\kappa_1}\right)+
p_{1l}\sin\left( \frac{p_{2\gamma}}{\kappa_1}\right) - \sin\left( \frac{p_{1\gamma}}{\kappa_1}\right)p_{2l} = 0\;,
$$
cancels all of them. Consequently, we get the formulas (\ref{symp2}), (\ref{symp3}) and (\ref{symsp2}), (\ref{symsp3}) respectively. In the similar way we proceed
with rules (\ref{ten}) and (\ref{symsen}).}
\begin{eqnarray}
&{\bf i)}&p_{3\gamma} = p_{1\gamma} +p_{2\gamma}\;, \label{symp1}\\
&~~&p_{3k} = p_{1k} + p_{2k} + \left[\cos\left(\frac{ p_{1\gamma}}{\kappa_1}  \right)-1\right]p_{2k}+ p_{1k}\left[\cos\left(\frac{ p_{2\gamma}}{\kappa_1}  \right)-1\right]
\;, \label{symp2}
\end{eqnarray}
\begin{eqnarray}
&&p_{3l} =p_{1l} + p_{2l} +\left[\cos\left(\frac{ p_{1\gamma}}{\kappa_1}\right)-1\right]p_{2l} + p_{1l}\left[\cos\left(\frac{ p_{2\gamma}}{\kappa_1}  \right)-1\right]
\;, \label{symp3}\\
&&~~~~~~~~~~~~~~~~~~~~~~~~~~~~~~~~E_3 = E_1 \;+\; E_2\;, \label{symen}
\end{eqnarray}
\begin{eqnarray}
&{\bf ii)}&p_{3\gamma} = p_{1\gamma} + p_{2\gamma}\;, \label{symsp1}\\
&~~&p_{3k} = p_{1k}+p_{2k}+\left[\cos\left(\frac{E_1}{\kappa_2} \right)-1\right]p_{2k}+ p_{1k}\left[\cos\left(\frac{E_2}{\kappa_2} \right)-1\right]
\;, \label{symsp2}
\end{eqnarray}
\begin{eqnarray}
&&p_{3l} = p_{1l} + p_{2l}+\left[\cos\left(\frac{E_1}{\kappa_2} \right)-1\right]p_{2l}+p_{1l}\left[\cos\left(\frac{E_2}{\kappa_2} \right)-1\right]
\;, \label{symsp3}\\
&&~~~~~~~~~~~~~~~~~~~~~~~~~~~~~~~~E_3 = E_1\; +\; E_2\;, \label{symsen}
\end{eqnarray}
\begin{eqnarray}
&{\bf iii)}&p_{3\gamma} \;=\; p_{1\gamma} + p_{2\gamma}\;\;,\;\;
p_{3k} \;=\; p_{1k} + p_{2k}\;\;,\;\;
p_{3l} \;=\; p_{1l} + p_{2l}\;\;,\;\; E_3 \;=\; E_1 \;+ \;E_2 \;~. \nonumber
\end{eqnarray}
Then, the energy-momentum conservation law for two-particle system takes the form
\begin{eqnarray}
&&\bar{p}_1 +_{\kappa_i} \bar{p}_2 = \bar{p}_{{\rm tot}} ={\rm const.}\;, \label{lielaw1}\\
&&E_1 +_{\kappa_i} E_2 = E_{{\rm tot}} ={\rm const.}\;, \label{lielaw2}
\end{eqnarray}
with the components of total vector $\bar{p}_{{\rm tot}}$ and with total energy $E_{{\rm tot}}$ given by equations (\ref{p1})-(\ref{ten}) in the case of twisted coproduct $\Delta(P)$ as well as by {\bf i)}, {\bf ii)} and {\bf iii)} formulas for their symmetrized counterparts respectively.

Let us now turn to the special situation when $\bar{p}_{{\rm tot}} = 0$ and $E_{{\rm tot}} = 0$, i.e., when
\begin{eqnarray}
\bar{p}_1 +_{\kappa_i} \bar{p}_2 = 0\;, \label{van1}
\end{eqnarray}
and
\begin{eqnarray}
E_1 +_{\kappa_i} E_2 = 0\;, \label{van2}
\end{eqnarray}
or, equivalently
\begin{eqnarray}
p_{3\gamma} = 0 = p_{3k} = p_{3l}\;\;\;,\;\;\;E_3 = 0\;. \label{equivalent}
\end{eqnarray}
Then, by direct calculation one can check that the conditions (\ref{equivalent}) are satisfied by classical set of solutions (\ref{forem1}), (\ref{forem2}) for all three Lie-algebraically
deformed Hopf algebras ${\cal U}_{\kappa_i}({\cal G})$ for both symmetrized and nonsymmetrized addition rules, as well as by
\begin{eqnarray}
&&p_{1\gamma} = - p_{2\gamma} = \kappa_1\pi\left(n-\frac{1}{2}\right)\;\;;\;\;n\in \mathbf{Z}\;,\label{solutionq1a} \\
&&E_1 = - E_2\;, \label{solutionq1b}\\
&&p_{1k}\;\;{\rm and}\;\;p_{1l} - {\rm arbitrary\;real\;numbers}\label{solutionq1c}\;,
\end{eqnarray}
in the case of the first quantum group ${\cal U}_{\kappa_1}({\cal G})$ and
\begin{eqnarray}
&&p_{1\gamma} = - p_{2\gamma}\;,\label{solutionq2a} \\
&&E_1 = - E_2 = \kappa_2\pi\left(n-\frac{1}{2}\right)\;\;;\;\;n\in \mathbf{Z}\;, \label{solutionq2b}\\
&&p_{1k}\;\;{\rm and}\;\;p_{1l} - {\rm arbitrary\;real\;numbers}\label{solutionq2c}\;,
\end{eqnarray}
for the second Hopf structure ${\cal U}_{\kappa_2}({\cal G})$ only after the symmetrization. Obviously, for parameters $\kappa_i$ approaching infinity the above formulas become commutative.

\section{Final remarks}

In this article we provide the addition rules for two momenta of particles in the case of three Lie-algebraically and two canonically twist-deformed Galilei Hopf algebras. The proposed
prescription remains consistent with $\Delta(P)$-coproducts given by the formulas (\ref{dlww3v}), (\ref{zcoppy1}), (\ref{ffdlww2.2}), (\ref{america1}), (\ref{sffdlww2.2}), (\ref{america2}),
(\ref{coppy1}) and (\ref{cop0}) respectively. Besides, we formulate the energy-momentum conservation principle for all considered systems. Particularly, we show that the total energy-momentum
vanishing condition $\bar{p}_{\rm tot}=0$ and $E_{\rm tot}=0$ leads to the quantization of three-momentum in the case of ${\cal U}_{\kappa_1}({\cal G})$ Hopf algebra as well as
to the discretisation of energy values for ${\cal U}_{\kappa_2}({\cal G})$ Hopf structure only in the case of the symmetrized addition rules {\bf i)}, {\bf ii)} and {\bf iii)}.

It should be noted that the above considerations can be extended to the N-particle twist-deformed relativistic and nonrelativistic kinematical models as well. The works in this direction
already started and are in progress.

\section*{Acknowledgments}

The author would like to thank J. Lukierski for valuable discussions.

\end{document}